# Exciton density waves in Coulomb-coupled dual moiré lattices

Yihang Zeng[1,*], Zhengchao Xia[2,*], Roei Dery[1], Kenji Watanabe[3], Takashi Taniguchi[3], Jie Shan[1,2,4,**], Kin Fai Mak[1,2,4,**]

[1]Department of Physics, Cornell University, Ithaca, NY, USA.
[2]School of Applied and Engineering Physics, Cornell University, Ithaca, NY, USA.
[3]National Institute for Materials Science, 1-1 Namiki, 305-0044 Tsukuba, Japan.
[4]Kavli Institute at Cornell for Nanoscale Science, Ithaca, NY, USA.
[*]These authors contributed equally.
[**]Email: jie.shan@cornell.edu; kinfai.mak@cornell.edu

**Strongly correlated bosons in a lattice are a platform to realize rich bosonic states of matter and quantum phase transitions [1]. While strongly correlated bosons in a lattice have been studied in cold-atom experiments [2-4], their realization in a solid-state system has remained challenging [5]. Here we trap interlayer excitons--bosons composed of bound electron-hole pairs--in a lattice provided by an angle-aligned $WS_2$/bilayer $WSe_2$/$WS_2$ multilayer; the heterostructure supports Coulomb-coupled triangular moiré lattices of nearly identical period at the top and bottom interfaces. We observe correlated insulating states when the combined electron filling factor of the two lattices, with arbitrary partitions, equals to $\frac{1}{3}, \frac{2}{3}, \frac{4}{3}$ and $\frac{5}{3}$. These new states can be interpreted as exciton density waves in a Bose-Fermi mixture of excitons and holes [6, 7]. Because of the strong repulsive interactions between the constituents, the holes form robust generalized Wigner crystals [8-11], which restrict the exciton fluid to channels that spontaneously break the translational symmetry of the lattice. Our results demonstrate that Coulomb-coupled moiré lattices are fertile ground for correlated many-boson phenomena [12, 13].**

Two-dimensional moiré materials have emerged as a highly controllable quantum system for exploring the effects of strong correlation and band topology [14-18]. A plethora of correlated states, including Mott insulators [10, 19-21] and generalized Wigner crystals [8-11], have been demonstrated for electrons in a moiré lattice formed in semiconducting transition metal dichalcogenide (TMD) bilayers. However, strongly correlated excitons are generally not achievable in a single moiré lattice [22, 23] because of the short exciton lifetime. Two symmetric moiré lattices separated by a thin barrier have been proposed to realize exciton fluids in a lattice [13] (Fig. 1a-c). Excitons here are the bound states of electrons in one lattice and empty sites ('holes') directly above (or below) in the other lattice; both are in the lowest moiré minibands formed from the conduction (or valence) bands of the host semiconductor. Strong binding is expected when the electronic correlation is strong in each lattice, and the lattice separation, $d$, is small compared to the moiré period, $a_M$. Similar to the case of Coulomb-coupled electron-hole double layers without a lattice [24-32], the spatially separated double lattice structure significantly suppresses interlayer electron tunneling and lengthens the exciton lifetime to allow strongly correlated excitons. Many bosonic phases, including excitonic Mott insulators, Wigner solids, superfluids and supersolids, have been predicted [12, 13].

Here we demonstrate two Coulomb-coupled moiré lattices of nearly identical period in angle-aligned WS$_2$/bilayer WSe$_2$/WS$_2$ multilayers (Fig. 1b,c). In contrast to the recently studied excitons in a moiré lattice Coulomb-coupled to a lattice-free monolayer [33, 34], both electrons and holes here are in the flat moiré minibands, and substantially stronger correlations are expected. We observe new insulating states at combined electron filling factor of the two lattices, $\nu = \frac{1}{3}, \frac{2}{3}, \frac{4}{3}, \frac{5}{3}$, in addition to an exciton fluid at $\nu = 1$ that has been recently reported in the coupled moiré lattice-monolayer structure [33, 34]. These new states are distinct from the generalized Wigner crystals in a single moiré lattice [8-11] in that a fluid of excitons is present. They provide a rare realization of exciton density waves with exciton density spontaneously breaking the translational symmetry of the lattice [6, 7, 35, 36]. The Coulomb-coupled moiré lattices provide a promising platform to realize other exotic bosonic phases and quantum phase transitions between them.

Figure 1c illustrates a dual-gated device of an angle-aligned WS$_2$/bilayer WSe$_2$/WS$_2$ multilayer. Triangular moiré lattices of period $a_M \approx 8$ nm are formed at both the top and bottom WS$_2$/WSe$_2$ interfaces because the two materials have a 4% lattice mismatch [10, 19]. For small twist angle, the moiré period is insensitive to small variations in the twist angle, thus enabling the creation of two moiré lattices of nearly identical period. We focus on the case of electron doping to maximize the correlation effects since the conduction minibands are flatter in this system [11]. Figure 1b illustrates the alignment of the conduction minibands. For the relevant doping range, electrons are located in the WS$_2$ layers. The WSe$_2$ bilayer separates the moiré lattices by $d \approx 2$ nm (center-to-center distance) and acts as a tunnel barrier with a barrier height around 250 meV [19]. To further suppress the electronic coupling between the two lattices, we align the two WS$_2$ monolayers at 180°; interlayer electron tunneling is spin-forbidden due to spin-valley locking in each monolayer [37]. Figure 1c also shows that the coupled moiré lattices are grounded and encapsulated in a top and bottom gate made of hBN dielectrics and graphite gate electrodes. The two gates independently control the combined filling factor $\nu$ (or total doping density in units of moiré density) and the out-of-plane electric field $E$ in two moiré lattices. The latter tunes the relative band alignment and distribution of the electrons between the two lattices at a fixed $\nu$.

To probe the insulating states in the coupled moiré lattices, we employ an optical sensing technique [11] by placing a WSe$_2$ monolayer above the top WS$_2$ layer. A thin ($\approx$ 1-2 nm) hBN spacer is introduced to prevent direct electronic coupling between the sensor and the sample, but is sufficiently thin so that the charge compressibility or dielectric constant [38] of the sample can be probed through dielectric screening of the 2s exciton resonance of the WSe$_2$ sensor. Throughout the measurements, the sensor is kept charge neutral (the 2s exciton resonance is quenched in a doped sensor). Unless otherwise specified, all results are obtained at a temperature of 3.5 K. See Methods for details on the device fabrication and optical measurements.

Figure 1d-f show the reflectance contrast (RC) spectrum of the sensor 2s exciton as a function of gate voltage (lower axis) and total doping density (upper axis). The three examples correspond to electron doping solely in the bottom lattice (1d), in the top lattice

(1e), and evenly in both lattices (1f). The accessible doping range in these examples is limited (and also different) in order to keep the sensor charge neutral. An insulating state in the sample is identified when the 2s exciton resonance exhibits a blueshift and an enhanced spectral weight [11]. When the electrons are in one of the lattices solely (Fig. 1d, e), the observation is similar to the reported result in single WS$_2$/WSe$_2$ moiré structures. In particular, insulating states are observed at $\nu = \frac{1}{3}, \frac{2}{3}, 1, 2$. Compared to the case of a single moiré structure [11], here fewer correlated insulating states at fractional fillings (generalized Wigner crystals) can be identified; the weaker ones are presumably suppressed by the stronger dielectric screening and/or disorder effects in coupled moiré lattices. In addition, we observe an overall stronger renormalization of the 2s exciton resonance in Fig. 1e than in Fig. 1d. This is a manifestation of higher sensor sensitivity to the top lattice because of its closer proximity [39]. Using the gate voltages for the assigned insulating states and the gate capacitances from an independent measurement, we determine the moiré density to be $(1.98 \pm 0.10) \times 10^{12}$ and $(2.02 \pm 0.10) \times 10^{12}$ cm$^{-2}$ for the top and bottom lattices, respectively. The two lattice periods are therefore identical within the uncertainties.

When the electrons are doped evenly in both lattices by setting $E \approx 0$ V/nm (Fig. 1f), we observe additional insulating states at $\nu = \frac{4}{3}, \frac{5}{3}$ apart from $\nu = \frac{1}{3}, \frac{2}{3}, 1, 2$. These states are absent when the electrons are doped solely in one lattice because the second Hubbard band of the WS$_2$/WSe$_2$ moiré lattice is involved and the electronic correlations are substantially weaker [11]. This example illustrates the sensitivity of the correlated states to the charge configurations in both lattices as well as the sensitivity of the exciton sensor to these states.

We map the dielectric constant of the coupled moiré lattices as a function of $\nu$ and $E$ in Fig. 2a. We represent the dielectric constant by $R_{2s}$, the amplitude of the sensor 2s exciton RC (Extended Data Fig. 4), as done in a recent study [38]; enhanced $R_{2s}$ corresponds to a charge-incompressible (insulating) state. The regions in white cannot be accessed in the current device. The map is also more sensitive to the response of the top moiré lattice because it is closer to the sensor. As expected, the largest $R_{2s}$ is observed when the entire structure is charge-incompressible, for instance, when both lattices have integer fillings ($\nu_t$ and $\nu_b$ denote the filling factor of the top and bottom lattices, respectively). We also observe extended regions with enhanced $R_{2s}$ when the top lattice is charge-incompressible, and the bottom, generally charge-compressible. Three regions are identified and denoted in orange in Fig. 2b. They are assigned as $\nu_t = 0$, 1, and 2 with the chemical potential inside the semiconductor band gap, the Mott gap, and the first moiré band gap of the top lattice, respectively. The regions are extended because a finite electric field is required to Stark shift the minibands in the two lattices to overcome the finite charge gap of the top lattice [33, 40].

To complete the electrostatic phase diagram, we add in Fig. 2d the mirror image of the orange regions with respect to zero field (blue regions) to represent regions of gapped bottom moiré lattice ($\nu_b = 0$, 1 and 2). Similarly, the top lattice is generally charge-compressible in these regions. The sensor 2s exciton is, however, effectively screened by the top lattice and no longer sensitive to the compressibility of the bottom lattice. The electrostatic phase diagram of Fig. 2d is fully consistent with an independent measurement based on the moiré excitons in $WS_2$ (Fig. 2c). The RC of the fundamental moiré exciton in $WS_2$, $R_{MX}$, decreases significantly when at least one of the $WS_2$ layers is electron-doped [11] (Extended Data Fig. 3). The boundary between the high and low values of $R_{MX}$ agrees well with the combined boundary of the region of $\nu_t = 0$ and $\nu_b = 0$ (dashed lines). See Methods for additional discussions on the electrostatic phase diagram.

We now focus on the unfilled regions in the phase diagram, in which both lattices would generally be charge-compressible in the absence of strong correlation between the lattices. Figure 3a is a closer examination of the electronic compressibility in the first unfilled region with $0 \leq \nu_t, \nu_b \leq 1$ (enclosed by dashed lines). Remarkably, we identify incompressible states from the enhanced $R_{2s}$ at $\nu = \frac{1}{3}, \frac{2}{3}, 1, \frac{4}{3}, \frac{5}{3}$ for all electric fields in the given range; and the electric-field dependence of $R_{2s}$ is smooth for all states (Fig. 3b). Figure 3c shows a horizontal linecut of the phase diagram at $E \approx 0$ V/nm under varying temperatures. As temperature increases, the incompressible states at the fractional fillings disappear around 30-35 K; but the $\nu = 1$ state persists up to much higher temperatures. These states are non-topological from the magnetic-field dependence study in Extended Data Fig. 5.

The insulating states outside the dashed box in Fig. 3a are well understood, with each lattice in an insulating state (independent of the other). For instance, the $\nu = 1$ state corresponds to $(\nu_t, \nu_b) = (0, 1)$ under large upward electric fields and $(1, 0)$ under large downward fields, that is, one lattice is unfilled and the other is a Mott insulator. The $\nu = \frac{1}{3}$ state corresponds to $(\nu_t, \nu_b) = (0, 1/3)$ and $(1/3, 0)$, that is, one lattice is unfilled and the other is a generalized Wigner crystal.

The nature of the insulating states inside the dashed box is completely different. Here the electrons are continuously transferred from one lattice to the other by the electric field. They arise from inter-lattice (or inter-layer) electronic correlations and can be viewed as excitonic insulators [12, 13, 33, 34]. We first discuss the case of $\nu = 1$ with one lattice unfilled and the other being a Mott insulator. Electrons are transferred from the Mott insulator to the empty lattice but remain bound to the empty sites in the original lattice by inter-lattice 'onsite' Coulomb repulsions (onsite Coulomb repulsions exceed other energy scales in the problem). These bound states can be viewed as interlayer excitons, charge-neutral particles that can hop around the lattice (Fig. 4a). The result is an exciton fluid in a lattice with a dipole-dipole repulsion of $\sim (V - V')$. Here $V$ and $V'$ ($\lesssim V$ in the limit $d \ll a_M$) denote the long-range Coulomb repulsion in the same lattice and between the lattices, respectively. The smooth electric-field dependence in Fig. 3b shows that the excitonic insulator is exciton-compressible. The ground state of such an exciton fluid is expected to be a superfluid in the weak-disorder limit [12, 13].

At total fractional fillings, we start with an empty lattice and a generalized Wigner crystal in the other. (We illustrate the cases of $\nu = \frac{1}{3}, \frac{2}{3}$ in Fig. 4b, c; the physics of $\nu = \frac{4}{3}, \frac{5}{3}$ is nearly identical.) Electrons are transferred from the generalized Wigner crystal to the empty lattice but remain bound to the empty sites in the original lattice by the inter-lattice (or inter-layer) long-range Coulomb repulsion $V'$. In order to minimize the total intra- and inter-lattice Coulomb repulsions ($V$ and $V'$), the electrons from both layers combine to form an "inter-layer" Wigner crystal, which defines channels in a lattice that guide the hopping of the interlayer excitons (Fig. 4b,c); the exciton-exciton interaction is $V - V'$ (as in the case for $\nu = 1$). In comparison to the exciton fluid at $\nu = 1$, the excitonic insulators at total fractional fillings are also exciton-compressible (Fig. 3b). But the exciton density distribution spontaneously breaks the translational symmetry of the lattice. In the limit of $d \ll a_M$, the melting temperature of these exciton density waves is expected to be similar to that of the generalized Wigner crystals [8, 10, 11], which is consistent with the experimental data in Fig. 3c. We note that exciton-electron phase separation into macroscopic domains is unstable because of the large $V \geq V'$.

In summary, we have observed new correlated insulating states at total fractional fillings $\frac{1}{3}, \frac{2}{3}, \frac{4}{3}, \frac{5}{3}$ (also $\frac{1}{2}$ in some devices) in two Coulomb-coupled moiré lattices. The observation of these states requires strong electronic correlation within each lattice and between the two lattices that are closely spaced ($d \ll a_M$), but electronically decoupled. Perfect alignment of the two lattices is not required as long as the relative displacements are uniform and below the size of the electron Wannier functions (~ 2-3 nm) in the moiré lattice (see Methods for more discussions). The coupled moiré lattices demonstrated here open new doors to the search of exotic many-boson phenomena. In particular, the observed correlated insulating states can be viewed as exciton density waves emerged in an exciton-hole Bose-Fermi mixture [6, 7] (see Methods for discussions in a particle-hole transformation picture). In the weak-disorder limit, they are expected to possess finite superfluid densities in the ground state and are therefore supersolids, as predicted by theoretical studies on atomic Bose-Fermi mixtures [6, 7]. These states provide a route to realize exciton supersolidity through demonstration of spontaneous exciton phase coherence in future studies.

**Methods**
**Device fabrication and operation**
The devices (Fig. 1c) were fabricated using the layer-by-layer dry transfer method described in earlier studies [11, 19]. In short, flakes of monolayer $WS_2$, monolayer and bilayer $WSe_2$, few-layer hBN, and few-layer graphite were exfoliated from bulk crystals onto silicon substrates with a 285-nm oxide layer. They were identified by the reflectance contrast under an optical microscope and stacked in the desired sequence. The finished stack was transferred onto pre-patterned Au electrodes on silicon substrates. In the $WS_2$/2L-$WSe_2$/$WS_2$ moiré structure, one $WS_2$ monolayer is angle-aligned, and the other is anti-aligned, with the $WSe_2$ bilayer. The two $WS_2$ monolayers were cut from the same monolayer flake using an atomic force microscope (AFM) tip. Optical second-harmonic generation (SHG) was used to determine the crystal orientations prior to transfer. The monolayer $WSe_2$ sensor was placed above the moiré structure with a 1-2 nm hBN spacer and is not angle-aligned with the moiré structure. The sensor and the moiré structure are grounded by the same few-layer graphite electrode. For the device shown in the main text, the thickness of the hBN gate dielectric for the top and bottom gates is determined by AFM to be $d_{tg} = 25\pm0.3$ nm and $d_{bg} = 28\pm0.3$ nm, respectively. The top and bottom gate voltages ($V_{tg}$, $V_{bg}$) independently tune the out-of-plane electric field and the doping density in the moiré structure. The field, $E = (V_{tg}/d_{tg} - V_{bg}/d_{bg})/2$, is defined positive when it points from the top to the bottom moiré lattice. The normalized gate voltage, $V_g = V_{tg} + (d_{tg}/d_{bg})V_{bg}$, is proportional to the doping density and is used in Fig. 1f.

**Optical measurements**
The optical reflectance contrast (RC) was measured with the devices in a closed-cycle optical cryostat (attoDRY1000) down to 3.5 K. A tungsten halogen lamp was used as the light source and the incident power on the devices was kept below 0.8 nW. The spectrum of the reflected light from the devices was collected. The RC spectrum is defined as *(I-$I_0$)/$I_0$*, where $I_0$ is the reference spectrum and $I$ is the signal spectrum for a fixed doping density and out-of-plane electric field in the moiré structure. To obtain RC near the 2s exciton resonance of the sensor, we kept the sensor charge neutral; we used the spectrum measured with a heavily electron-doped sensor as the reference, for which the 2s exciton is quenched. To obtain RC near the $WS_2$ moiré exciton resonances, we used the spectrum measured with a heavily electron-doped moiré structure as the reference, for which the moiré exciton resonance is nearly quenched. Extended Data Fig. 3 shows the RC spectrum of device S1, including both the 1s and 2s exciton resonances in the $WSe_2$ sensor and the moiré excitons in $WS_2$ as a function of gate voltages. The $WSe_2$ moiré exciton resonances from the moiré structure have lower energies and are outside the spectral window. Extended Data Fig. 4 illustrates several horizontal linecuts from Extended Data Fig. 3 centered on the sensor 2s exciton and the $WS_2$ moiré exciton. Multilayer thin film analysis is required to describe the detailed line shape. For simplicity, we use the peak-to-peak variation of the features, $R_{2s}$ and $R_{MX}$, to denote the spectral weight of the 2s and the moiré exciton resonances, respectively. The moiré exciton resonances are relatively broad and the contribution from the two different $WS_2$ layers cannot be spectrally resolved. The moiré exciton RC in Fig. 2c therefore includes contributions from both layers.

**Electrostatic phase diagram of two Coulomb-coupled moiré lattices**
The electrostatic phase diagram of Fig. 2b, d can be qualitatively understood as follows. Inside the three orange-shaded regions, electron filling in the top layer $\nu_t$ is fixed at 0, 1 and 2, respectively (Fig. 2b). We first consider the region with $\nu_t = 0$. At $\nu = 0$, the Fermi level is inside the large semiconductor band gap of the heterostructure. With increasing $\nu$, electrons start to dope into the bottom layer for an upward field ($E < 0$) (and into the top layer for a downward field ($E > 0$) for $\nu_b = 0$). At the boundary of the region (denoted by an orange line), the Fermi level is fixed at the conduction band edge of the top layer. As $\nu$ increases along this line, the Fermi level sweeps through the lower and upper Hubbard bands of the bottom layer. Vertical jumps in $E$ are observed for the Fermi level inside the Mott gap ($\nu = 1$) and the moiré band gap ($\nu = 2$) of the bottom layer. The jump size multiplied by the top-bottom layer separation corresponds to the charge gap size [33, 40]. With the Fermi level inside the lower or upper Hubbard band of the bottom layer, the boundary shows a linear dependence between $E$ and $\nu$, the slope of which is determined by the thermodynamic density of states [33, 40].

The two other regions ($v_t = 1$ and 2) can be understood similarly. In particular, the lower (upper) boundary for $v_t = 1$ corresponds to the reference point where the Fermi level is fixed to the lower (upper) Hubbard band maximum (minimum) of the top layer. These two reference boundaries trace through the lower and upper Hubbard bands of the bottom layer in a way similar to the reference boundary for $v_t = 0$. The vertical distance in electric field between these two boundaries is proportional to the Mott gap size of the top layer. The region $v_t = 2$ has similar interpretations except that the Mott gap of the top layer is replaced by the moiré band gap. Furthermore, because the WS$_2$/2L-WSe$_2$/WS$_2$ moiré structure is symmetric, the discussions above apply equally well to the blue-shaded regions with $v_b = 0$, 1 and 2. The orange- and blue-shaded regions are symmetric about the $E = 0$ line (Fig. 2d).

The white regions in Fig. 2d correspond to the Fermi level inside the Hubbard bands of both layers. The electrons are added to the system along the filling factor axis and are continuously transferred between the two layers along the electric field axis. The system is in general charge-compressible except at total fractional filling factors corresponding to the excitonic insulating states.

**Effects of disorder in two Coulomb-coupled moiré lattices**
Compared to single moiré lattices, the coupled moiré lattice system introduces a new type of disorder involving the random variation in the relative displacement between the two lattices over a length scale long compared to $a_M$. The observed correlated states require a uniform relative displacement between the two moiré lattices over a sizable fraction of the probed area (about 1 micron in diameter). The current fabrication method does not have control over the moiré lattice alignment. Random displacement between the two moiré lattices can also arise from defects, unintentional strain etc. in each moiré layer. These effects are expected to limit our experiment. Indeed, results from different regions of the same device (Extended Data Fig. 1) and from different devices (Extended Data Fig. 2) show substantial variations compared to the single moiré samples. Whereas the excitonic insulating state at total filling $v = 1$ is observed in all areas and all devices, the insulating states at fractional fillings is observed only in about 20% of the sample areas (but in multiple devices). Future efforts are required to improve the uniformity of the coupled moiré lattice system.

**Particle-hole transformation picture**
The exciton density waves at total fractional fillings can also be viewed from a particle-hole transformation picture by ignoring the role of spins. Unlike the case of $v = 1$, particle-hole transformation in the bottom lattice cannot remove all of the fermionic degrees of freedom at total fractional fillings (Extended Data Fig. 7b,c); there are excess holes (open blue circles) after the formation of interlayer excitons, resulting in a Bose-Fermi mixture of excitons and holes. In this picture, the exciton density waves are defined by the repulsive interactions between the constituents including, in descending order of the energy scale, $V$ between the holes, $V'$ between the excitons and holes, and $V - V'$ between the excitons (as in the case for $v = 1$). In the strong correlation limit, where the interactions far exceed the hopping amplitude, the holes spontaneously form generalized Wigner crystals. The exciton-hole repulsion then guides the excitons to the channels defined by the hole Wigner crystals. A fluid of excitons that spontaneously break the translational symmetry of the lattice is formed.

**Acknowledgements**
We thank Allan MacDonald, Ya-Hui Zhang, Ashvin Vishwanath and Erich Mueller for fruitful discussions.

**Figures**

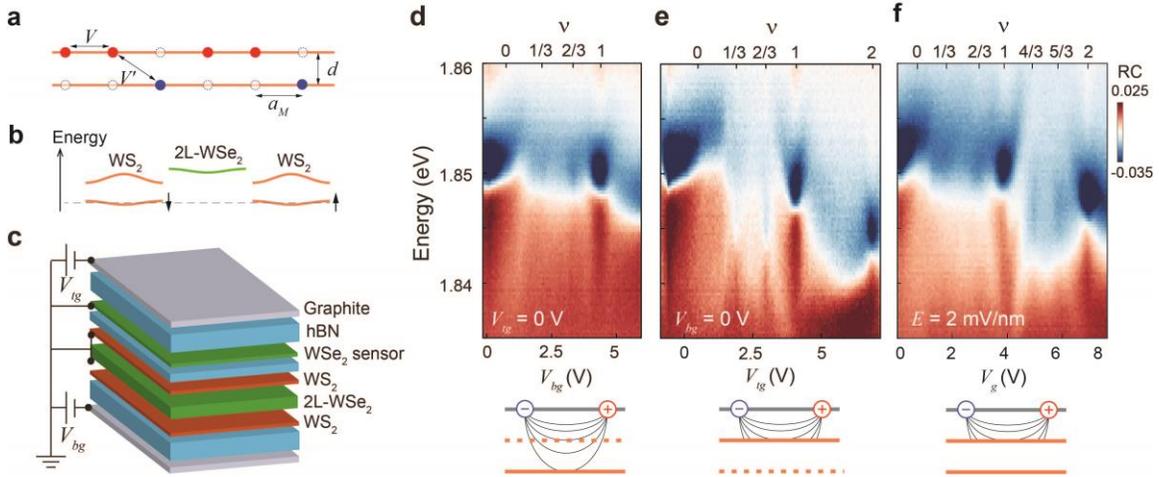

**Figure 1 | Coulomb-coupled moiré lattices. a,** Formation of interlayer excitons in two symmetric moiré lattices of period $a_M$ and separation $d$. Excitons are the bound states of electrons in one lattice (blue circles) and holes (empty circles) directly above it in the other lattice. **b,c,** Schematic of the conduction band alignment (**b**) and a dual-gated device (**c**) of an angle-aligned $WS_2$/bilayer $WSe_2$/$WS_2$ multilayer with two $WS_2$ monolayers aligned at 180°. Moiré lattices of period 8 nm are formed at the top and bottom $WS_2$/$WSe_2$ interfaces. They are separated by 2 nm. A $WSe_2$ monolayer sensor is separated from the top $WS_2$ layer by 4-5 layer hBN. The dashed line in **b** denotes the Fermi level under zero perpendicular electric field. The lowest moiré minibands have opposite spins (denoted by arrows) in the same valley. **d-f,** Top: Reflectance contrast spectrum of the sensor 2s exciton as a function of gate voltage (bottom axis) and total filling factor (top axis). Bottom: schematic of electron distribution in two lattices and their response to excitons in the sensor (grey). The solid and dashed orange lines denote a doped and empty lattice, respectively. Electrons are in the bottom lattice solely (**d**), the top lattice solely (**e**), and nearly equally in both lattices (**f**). The gate voltage in **f**, $V_g = V_{tg} + 0.88\ V_{bg}$, is normalized by the (different) values of the two gate capacitances.

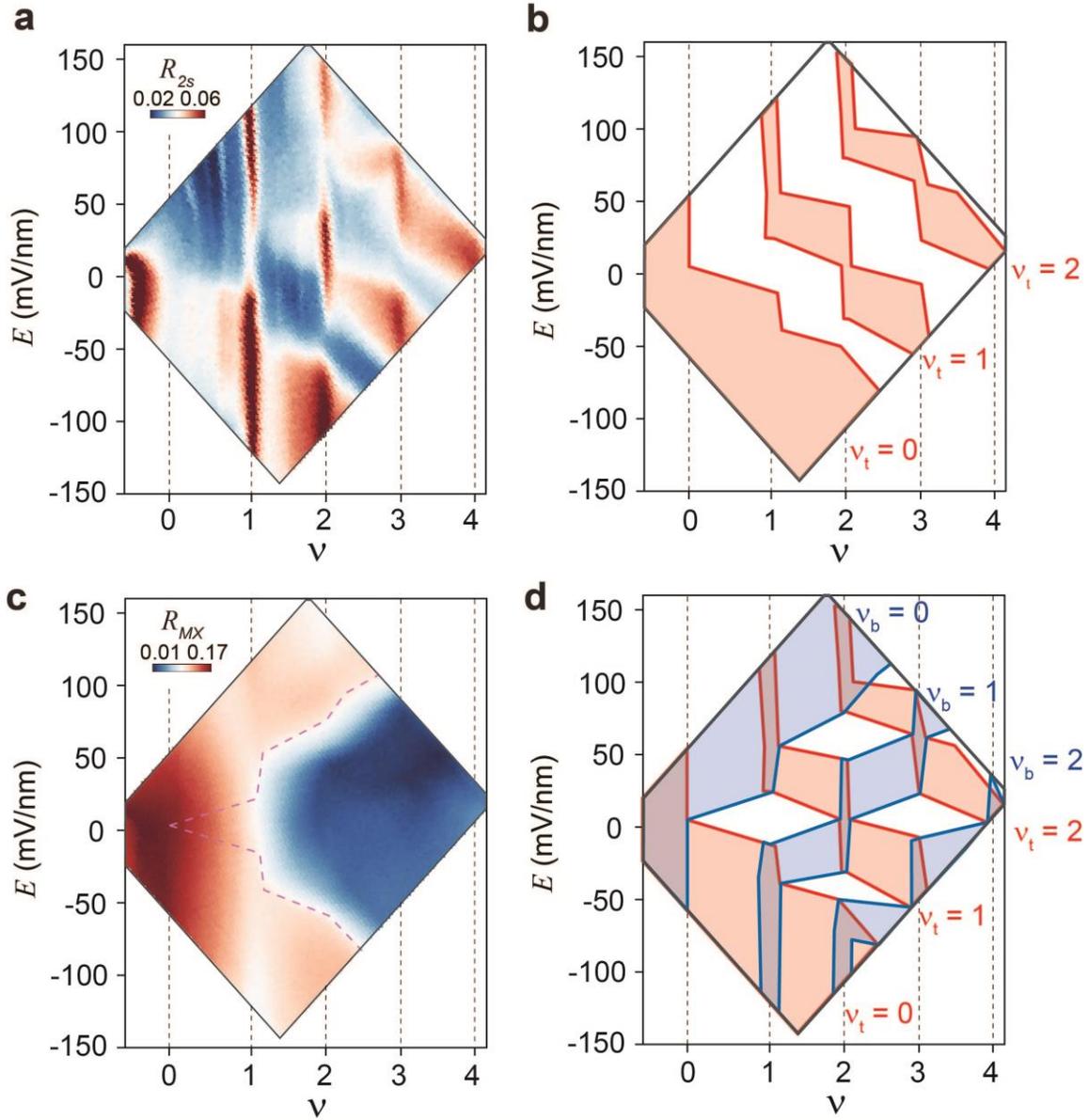

**Figure 2 | Electrostatic phase diagram of coupled moiré lattices. a,c,** Reflectance contrast amplitude of the sensor 2s exciton $R_{2s}$ (**a**) and the WS$_2$ moiré exciton $R_{MX}$ (**c**) as a function of total filling factor and electric field. **b,** Regions of gapped top lattice with top lattice filling $\nu_t = 0, 1, 2$ (orange-shaded). **d,** Electrostatic phase diagram including regions of gapped top lattice (orange-shaded) and bottom lattice (blue-shaded). The latter is a mirror image of the former. The dashed purple lines in **c** mark the onset of electron doping in both lattices.

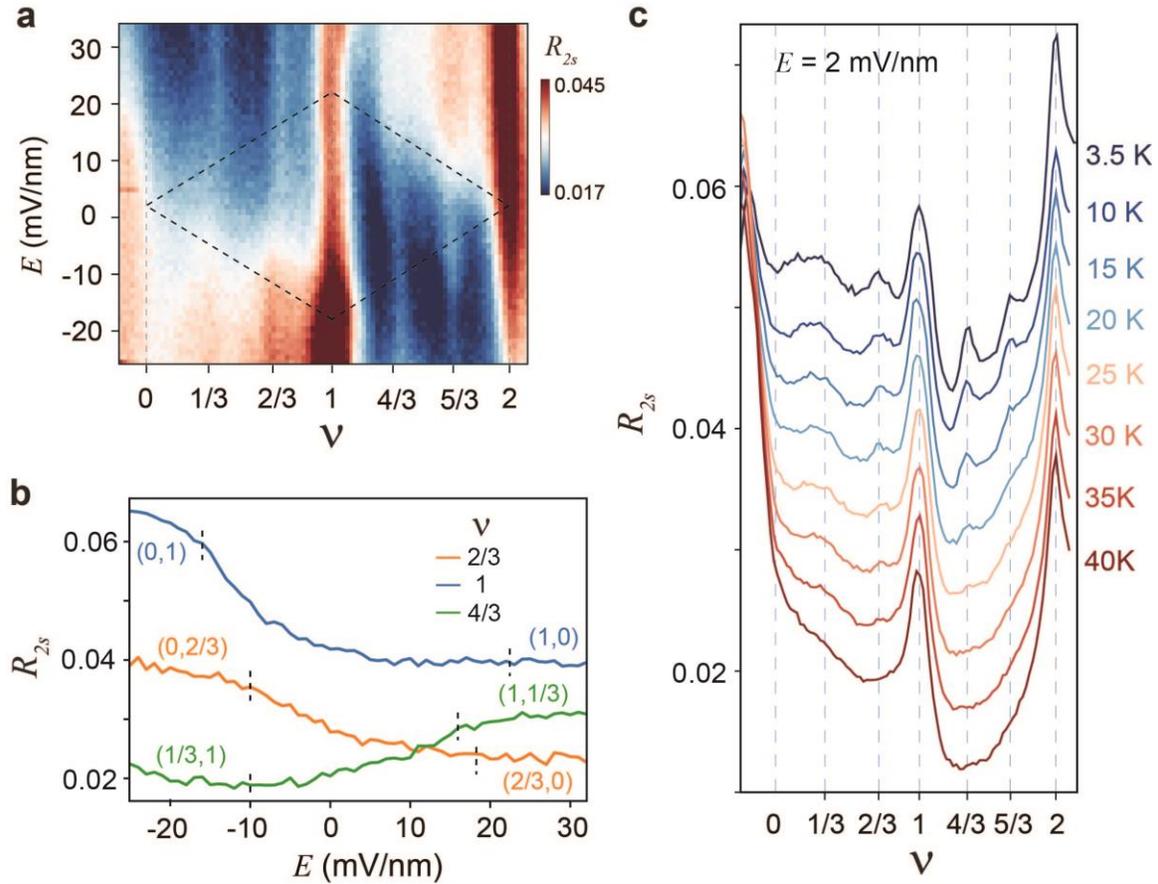

**Figure 3 | Correlated insulating states in coupled moiré lattices. a,** Electronic compressibility probed by the sensor 2s exciton as a function of total filling factor and electric field. Electrons are continuously transferred between the lattices inside the region enclosed by the dashed lines. Insulating states are identified from the enhanced $R_{2s}$ at $\nu = \frac{1}{3}, \frac{2}{3}, 1, \frac{4}{3}, \frac{5}{3}$. **b,** Vertical linecuts of **a** at $\nu = \frac{2}{3}$ (orange), $\nu = 1$ (blue) and $\nu = \frac{4}{3}$ (green). For each filling factor, two dashed marks show the boundary of the enclosed region in **a**. Outside the region, the electron filling factor in two lattices are given as $(\nu_t, \nu_b)$. **c,** Horizontal linecut of **a** near zero electric field at representative temperatures. The melting temperature is around 30 K for the $\nu = \frac{1}{3}, \frac{2}{3}$ states and around 20 K for the $\nu = \frac{4}{3}, \frac{5}{3}$ states.

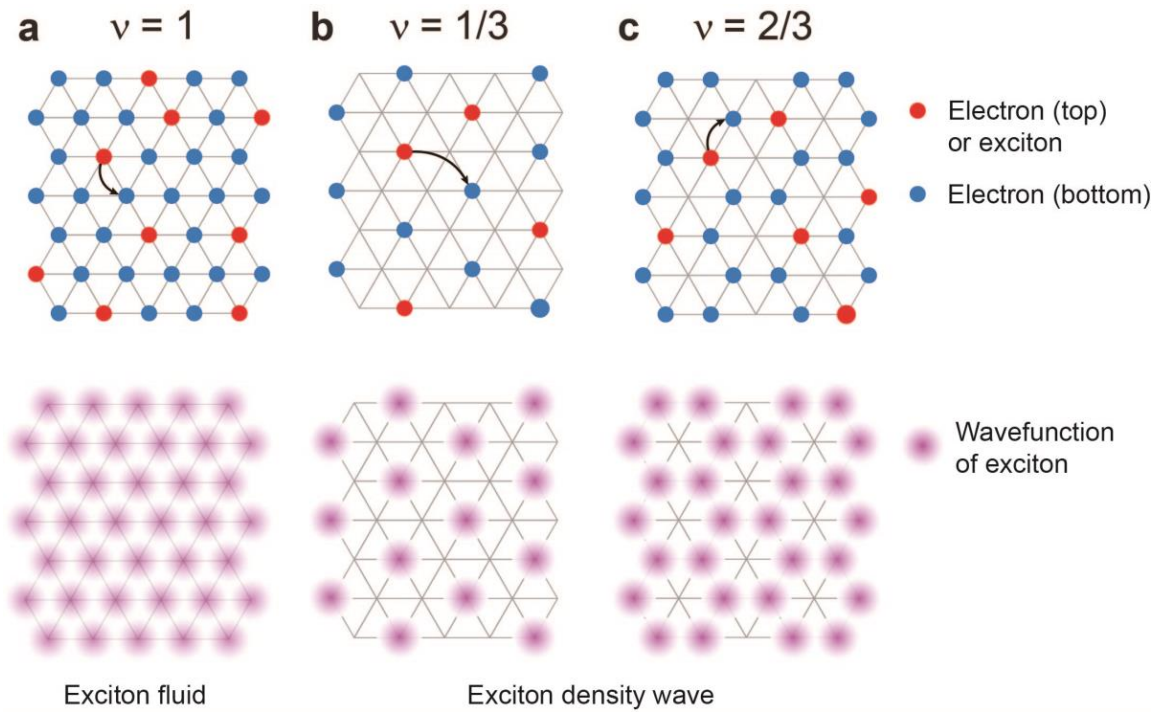

**Figure 4 | Exciton density waves. a-c,** Schematic representation of the inter-layer Mott insulator in coupled moiré lattices at total filling $\nu = 1$ (**a**), and of the inter-layer Wigner crystals at $\nu = \frac{1}{3}$ (**b**) and $\nu = \frac{2}{3}$ (**c**). Top, electrons in the top lattice (red) and bottom lattice (blue). Electrons in the top lattice are bound to the empty sites in the bottom lattice directly below them to minimize the Coulomb interactions. The bound states form interlayer excitons that can hop around the lattice (arrows). The exciton hopping is unrestricted in **a**, but is guided to the channels defined by the inter-layer Wigner crystals in **b** and **c**. Bottom: The exciton density distribution shows an exciton fluid at $\nu = 1$ and exciton density waves at $\nu = \frac{1}{3}, \frac{2}{3}$. The latter breaks the translational symmetry of the lattice.

**Extended data figures**

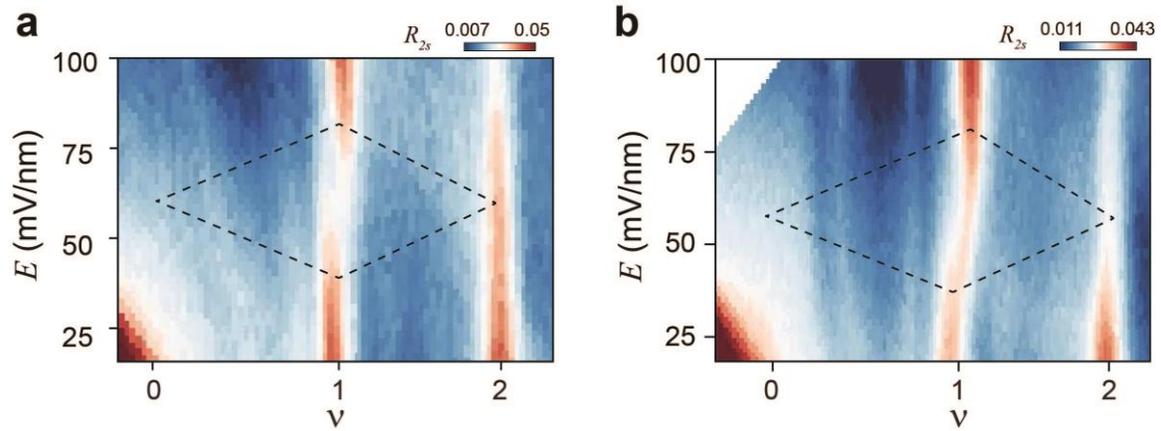

**Extended Data Figure 1 | Addition sample regions of device S1. a,b,** Dependence of $R_{2s}$ on the total filling factor and electric field at two additional regions of the same device as in the main figures. Electrons are in both moiré lattices in the region enclosed by the dashed lines. The insulating state at $\nu = 1$ is robust in all regions. Fewer and less robust insulating states are observed here particularly in **a**. The features appear curved at certain electric fields because the large contact resistance causes nonlinear gating effects. The electric field offset at $\nu_t = \nu_b$ is likely caused by the layer asymmetry in these regions.

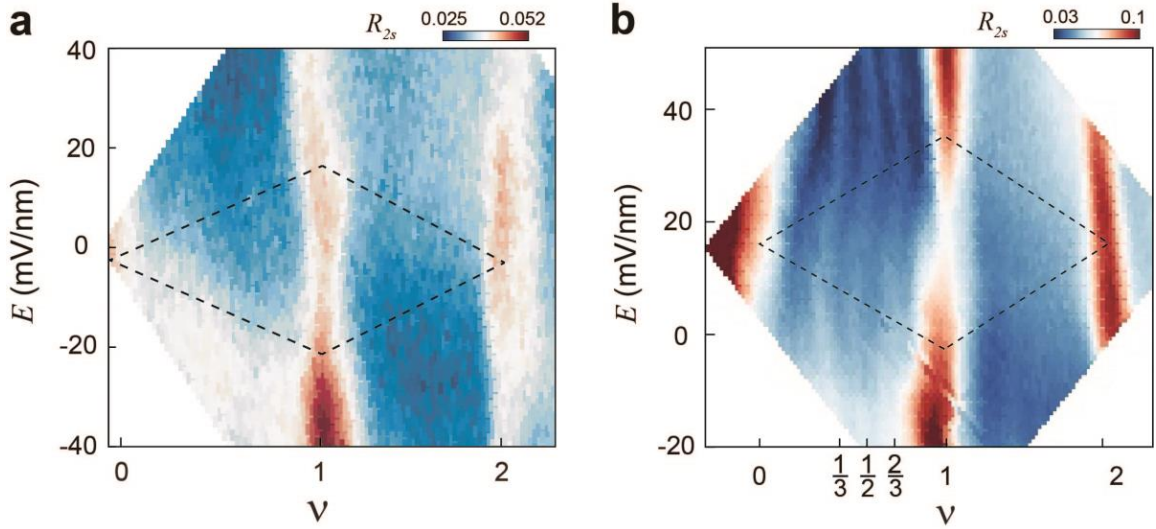

**Extended Data Figure 2 | Additional devices. a,b,** Dependence of $R_{2s}$ on the total filling factor and electric field for device S2 (**a**) and S3 (**b**). Electrons are in both moiré lattices in the region enclosed by the dashed lines. The insulating state at $\nu = 1$ is robust in both devices. The insulating states at total fractional fillings are observed only in device S3, including a new insulating state at $\nu = \frac{1}{2}$. The features appear curved at certain electric fields because the large contact resistance causes nonlinear gating effects.

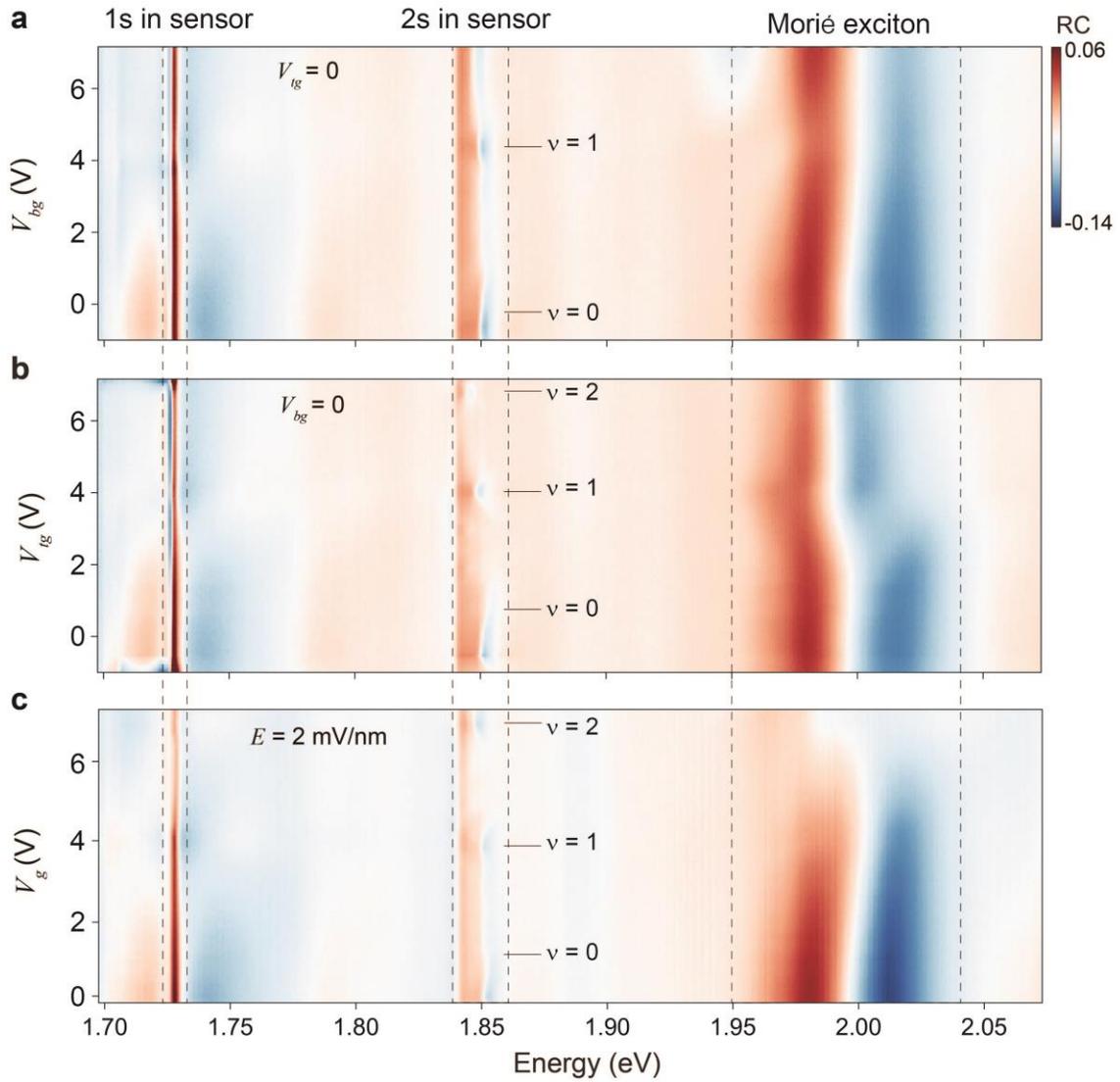

**Extended Data Figure 3 | Optical reflectance contrast spectrum of device S1. a-c,** Gate voltage dependence of the reflectance contrast spectrum covering the 1s and 2s resonances of the sensor and the $WS_2$ moiré excitons. Electrons are located in the bottom lattice solely (**a**), the top lattice solely (**b**) and equally in two lattices (**c**). The total filling factors $\nu = 0$, 1, and 2 are determined according to the sensor 2s exciton resonance.

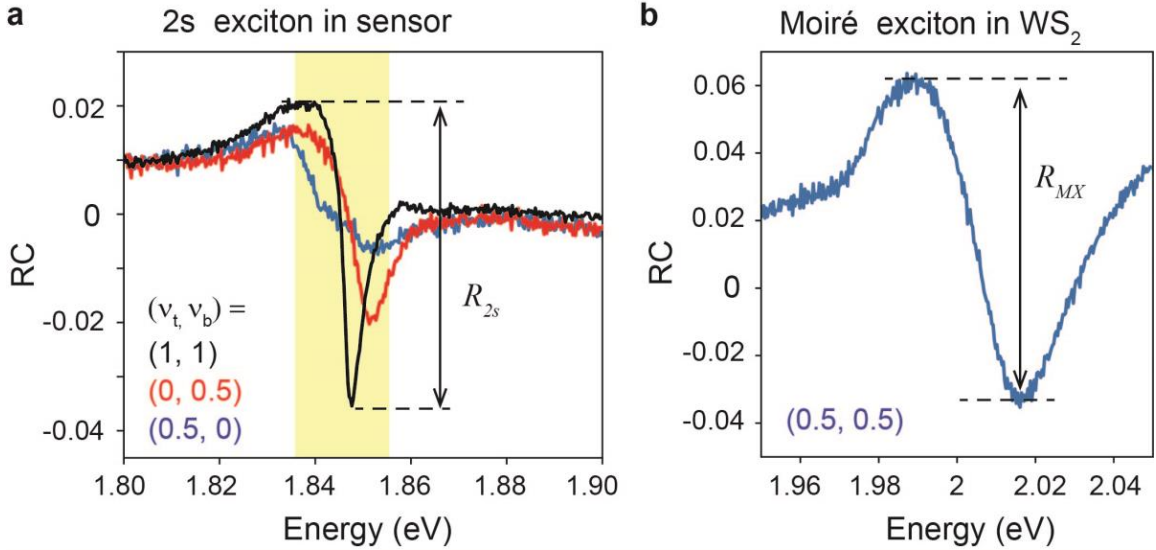

**Extended Data Figure 4 | Reflectance contrast (RC) spectra. a,b,** Representative RC spectra of the sensor 2s exciton (**a**) and the fundamental moiré exciton in $WS_2$ (**b**) of device S1 with lattice filling $(\nu_t, \nu_b)$. The spectral line shape is given by the optical interference effect in the multiple layer structure on the $Si/SiO_2$ substrate. We use the peak-to-peak amplitude of the RC between 1.836 and 1.853 eV ($R_{2s}$) as an indicator of the 2s exciton spectral weight (**a**), and between 1.95 and 2.05 eV ($R_{MX}$) for the moiré exciton (**b**).

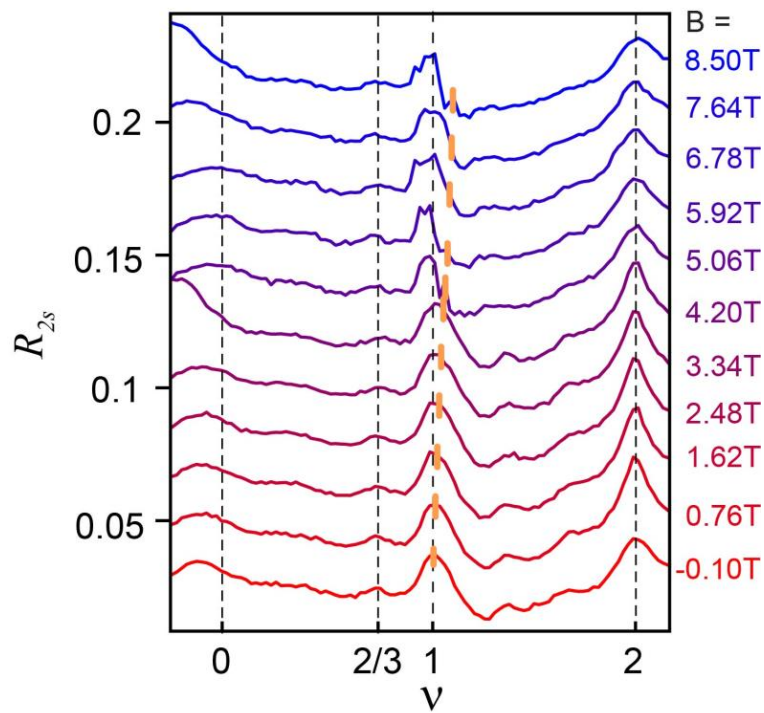

**Extended Data Figure 5 | Magnetic-field dependence.** Filling dependence of $R_{2s}$ at representative perpendicular magnetic fields. The curves are displaced vertically for clarity. The perpendicular electric field is 2 mV/nm. The black dashed lines mark total filling factor $\nu = 0$, 2/3, 1 and 2, at which insulating states are observed in the absence of the magnetic field. These states do not disperse with magnetic field and are therefore non-topological. The orange line at each magnetic field denotes the peak positions if the state were topological with Chern number 1.

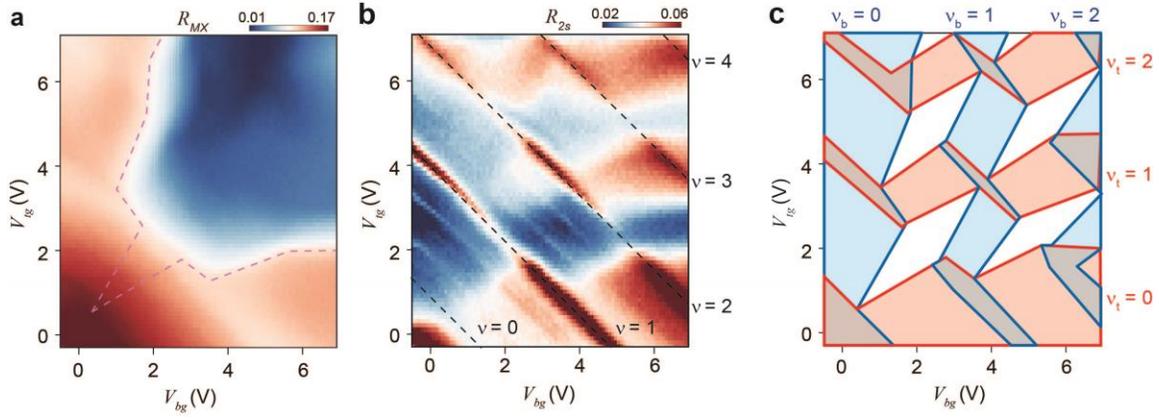

**Extended Data Figure 6 | Figure 2 in gate voltages.** Dependence of $R_{MX}$ (**a**) and $R_{2s}$ (**b**) on the gate voltages. Dashed lines in (**a**) mark the onset of electron doping in both lattices. Dashed lines in (**b**) mark the insulating states at total integer fillings $\nu = 0$, 1, 2, 3 and 4. **c,** The electrostatic phase diagram in gate voltages. The regions of gapped top lattice ($\nu_t = 0,1,2$) and bottom lattice ($\nu_b = 0,1,2$) are shaded in orange and blue, respectively.

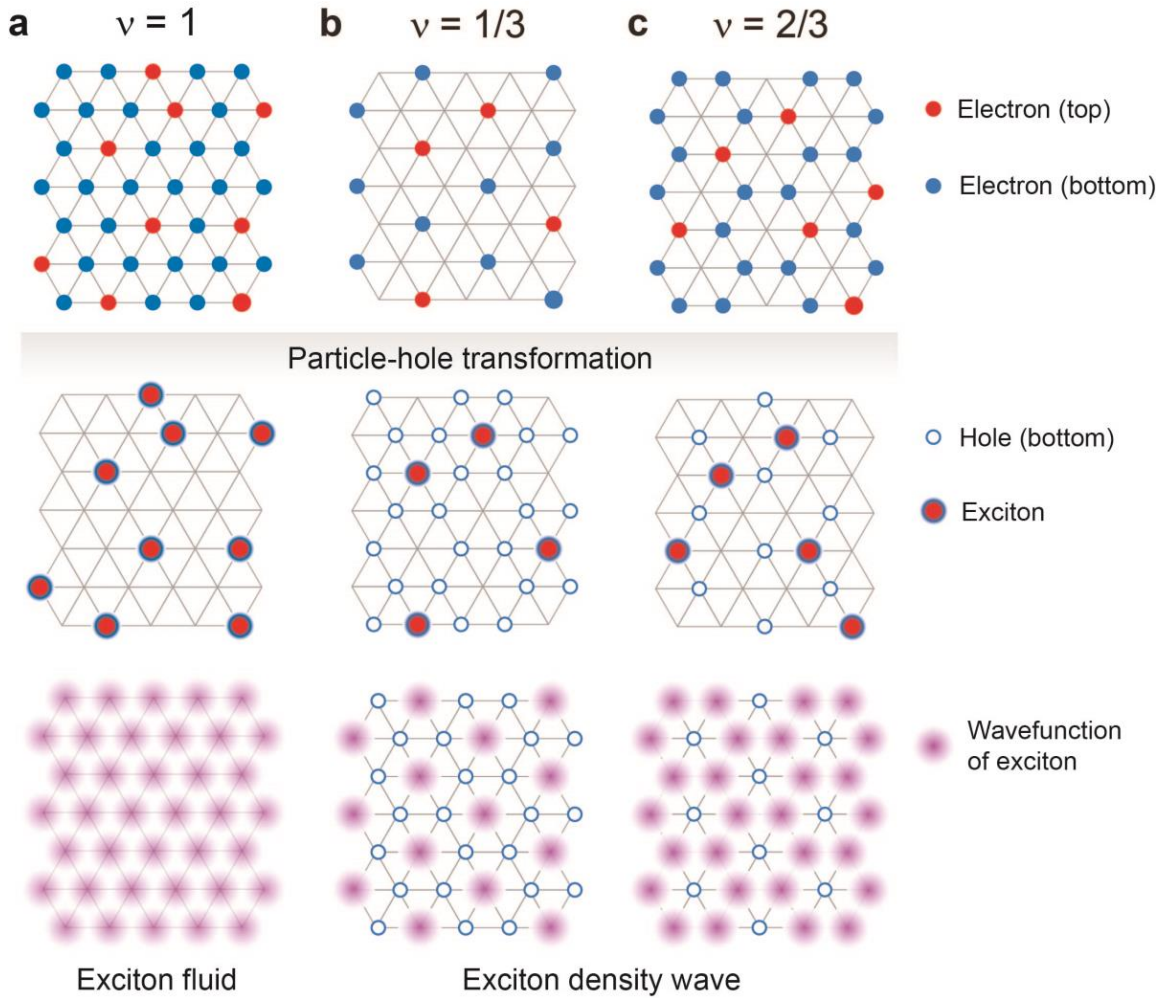

**Extended Data Figure 7 | Exciton density waves in a particle-hole transformation picture. a,-c,** Schematic representation of the correlated insulating states in coupled moiré lattices at total filling $\nu = 1$ (**a**), $\nu = \frac{1}{3}$ (**b**) and $\nu = \frac{2}{3}$ (**c**). Top, electrons in the top lattice (red) and bottom lattice (blue). Electrons in the top lattice are bound to the empty sites in the bottom lattice directly below them to minimize the Coulomb interactions. Middle, particle-hole transformation performed on the bottom lattice generates interlayer excitons (red-blue circles) and excess holes (empty blue circles). Bottom, the excess holes form generalized Wigner crystals. The excitons are guided to the channels defined by the hole Wigner crystals by the exciton-hole repulsion. The exciton density distribution shows an exciton fluid at $\nu = 1$ and exciton density waves at $\nu = \frac{1}{3}, \frac{2}{3}$. The latter breaks the translational symmetry of the lattice.